\documentclass{nature_figurecaption}

\usepackage{graphicx}
\usepackage[utf8x]{inputenc}
\usepackage{amsmath}
\usepackage{tikz}

\title{Bismuthene on a SiC Substrate: A Candidate for a New High-Temperature Quantum Spin Hall Paradigm}

\author{F. Reis$^{1}$*, G.~Li$^{2,3}$*, L.~Dudy$^1$,
  M.~Bauernfeind$^{1}$, S. Glass$^{1}$, W. Hanke$^3$, R.~Thomale$^3$, J. Sch\"afer$^1$, and R.~Claessen$^1$}

\begin{document}

\maketitle

\begin{affiliations}
 \item Physikalisches Institut and R\"ontgen Center for
   Complex Material Systems, Universit\"at W\"urzburg, D-97074
   W\"urzburg, Germany
\item Institute of Solid State Physics, Vienna University of Technology, A-1040 Vienna, Austria
 \item Institut f\"ur Theoretische Physik und Astrophysik, Universit\"at W\"urzburg, D-97074 W\"urzburg, Germany

* These authors contributed equally to this work.
\end{affiliations}

\begin{abstract}
Quantum spin Hall (QSH) materials promise revolutionary device applications
based on dissipationless propagation of spin currents. They are two-dimensional
(2D) representatives of the family of topological insulators, which exhibit
conduction channels at their edges inherently protected against scattering.
Initially predicted for graphene\cite{PhysRevLett.95.226801,PhysRevLett.95.146802},
and eventually realized in HgTe quantum wells\cite{Bernevig1757,Konig766}, in
the QSH systems realized so far\cite{Roth294,PhysRevLett.107.136603}, the
decisive bottleneck preventing applications is the small bulk energy gap of less than 30
meV, requiring cryogenic operation temperatures in order to suppress detrimental
bulk contributions to the edge conductance \cite{PhysRevX.3.021003,PhysRevLett.110.216402}.
Room-temperature functionalities, however, require much larger gaps. Here we
show how this can be achieved by making use of a new QSH paradigm based on
substrate-supported atomic monolayers of a high-$Z$ element. Experimentally, the material is
synthesized as honeycomb lattice of bismuth atoms, forming ''bismuthene``, on top of the
wide-gap substrate SiC(0001). Consistent with the theoretical expectations, the spectroscopic signatures in
experiment display a huge gap of $\approx 0.8\,$eV in bismuthene, as well as conductive
edge states. The analysis of the layer-substrate orbitals arrives at a QSH
phase, whose topological gap – as a hallmark mechanism – is driven directly by
the atomic spin-orbit coupling (SOC). Our results demonstrate how strained
artificial lattices of heavy atoms, in contact with an insulating substrate,
can be utilized to evoke a novel topological wide-gap scenario, where the
chemical potential is located well within the global system gap, ensuring pure edge
state conductance. We anticipate future experiments on topological signatures,
such as transport measurements that probe the QSH effect via quantized
universal conductance, notably at room temperature.
\end{abstract}

%
%
\indent While the QSH phase as such is characterized by universal topological
properties\cite{PhysRevLett.95.146802}, other aspects such as the bulk energy
gap are material-specific, and depend on the combination of SOC (in elements
of atomic number $Z$ scaling as $Z^4$), orbital hybridization, and the symmetries of
the layer-substrate system. In QSH honeycomb layers mimicking
graphene\cite{PhysRevLett.95.226801} (where minute SOC precludes a noticeable
gap), one obvious strategy to generate enhanced gaps is to replace C by heavier
atoms. For systems with the group-IV elements Si, Ge, and Sn, predicted gaps are
$2\,$meV\cite{PhysRevLett.107.076802}, $24\,$meV\cite{PhysRevLett.107.076802},
and $100\,$meV\cite{PhysRevLett.111.136804}, respectively. This applies to
hypothetical freestanding material – with poor chemical stability. Real-world
variants require a supporting substrate for epitaxial synthesis, which also
brings concurrent bonding interactions into play. Attempts were made to grow,
e.g., silicene on a metallic Ag substrate\cite{1.3524215,PhysRevLett.108.155501},
which however short-circuits any edge
states of interest. An insulating MoS$_2$ substrate can stabilize
germanene\cite{Zandvliet.PRL.2016}, yet the compressive strain renders the
material a metal. Stanene flakes on Bi$_2$Te$_3$ are plagued by the same
problem\cite{SCZhang.NatMat.2015}.\\
\indent Turning to group V of the Periodic Table, bismuth ($Z\,=\,83$) must be expected
to yield an even larger gap. It is predicted as topological non-trivial for thin free-standing
layers\cite{PhysRevLett.107.136805,PhysRevB.83.121310,PhysRevLett.97.236805}.
The conceptual existence of a monolayer bismuthene QSH phase on a silicon
substrate has been proposed\cite{zhou}, and extended to SiC by modelling of the
edge states\cite{bansil}. However, experimental growth of a hexagonal
single-layer Bi phase on Si(111) has failed\cite{failed1,0953-8984-15-17-302}.\\
%
%
\indent Our experimental realization of supported bismuthene employs a SiC(0001)
substrate, on which
we generate a ($\sqrt{3}\times\sqrt{3}$)R$30^\circ$ superstructure of Bi atoms
in honeycomb geometry, as illustrated in Fig. 1a (details in Extended Data Fig.
E1). The resulting lattice constant of $5.3\,$\AA$\,\,$ implies a sizeable tensile
strain of $+18\,\%$ compared to buckled Bi(111) bilayers. This causes a fully
planar configuration of the honeycomb rings. The synthesis (described in
Methods) starts from a hydrogen-etched SiC wafer on which a monolayer of Bi is
epitaxially deposited, giving rise to sharp electron diffraction peaks (Fig.
E2). A scanning tunneling microscopy (STM) overview (Fig. 1b) shows that the
whole surface is smoothly covered with flakes of typical diameter $\sim25\,$nm,
separated by phase-slip domain boundaries (related to the Bi-induced surface
reconstruction), and including occasional defects. Alternatively, the flakes are
also terminated by SiC substrate steps, see Fig. 1c. Inspection
of the layer on a smaller scale reveals the characteristic honeycomb pattern of
bismuthene (Fig. 1d). Detailed scrutiny of the honeycomb structure is provided
by the close-up STM in Fig. 1e. For both occupied and empty states,
respectively, the images show a honeycomb pattern throughout.\\
%
%
\indent To establish that the electronic structure of the synthesized material
corresponds to bismuthene, we performed angle-resolved photoelectron
spectroscopy (ARPES) and compared it with density-functional theory (DFT), see
Fig. 2. The electron bands in Fig. 2a are obtained using a hybrid
exchange-correlation functional, including SOC (see Methods). At the K-point,
the direct energy gap amounts to $1.06\,$eV (six orders of magnitude larger than
in graphene with $\sim 1\,\mu$eV\cite{PhysRevB.75.041401}). The conduction band
minimum at $\Gamma$ leads to an indirect gap of $0.67\,$eV in DFT. The
k-resolved dispersion from ARPES in Fig. 2b exhibits the characteristic maximum
at the K-point and a band splitting there, in close agreement with the DFT
overlay. From the ARPES close-up in Fig. 2c of the valence band maximum at K, we
derive a large band splitting of $\sim 0.43\,$eV. The constant energy maps in
Fig.~2d reflect the energy degeneracy of the K- and K'-points and give further proof
of the high degree of long-range order in the Bi honeycomb lattice.
The excellent agreement of ARPES and DFT results, thus, confirms the realization of a single
bismuthene layer on SiC. \\
%
%
\indent Next, we disentangle the key mechanisms that determine the energetics near the
Fermi level $E_\text{F}$ by reducing (downfolding) the full DFT band structure
to the relevant low-energy model (see Supplementary Information for details).
Inclusion of the \textit{substrate bonding} proves indispensable in this procedure. We decompose
the band structure into $\sigma$-bond contributions formed by Bi $6s$-, $p_x$- and
$p_y$-orbitals, and $\pi$-bond contributions formed by $p_z$-orbitals.
As a consequence of hybridization with the substrate via $\pi$-bonds, the
orbital content at low energies is predominantly of $\sigma$-type (Fig. 3a).
This manifests ''orbital filtering``, where the $p_x$- and $p_y$- orbitals play
the pivotal role close to $E_\text{F}$, while the $p_z$-orbital is projected out
of this energy region.\\
\indent Focussing on the low-energy bands around the Fermi level, we find that without
relativistic effects the mere orbital hybridization in the $\sigma$-bonding
sector (with eight bands due to two-fold orbital, spin, and sublattice degrees of freedom)
yields a Dirac-like band crossing at the K point, see Fig. 3b. SOC comes into play via two
leading contributions: First, the \textit{local (on-site) SOC term} $\lambda_\text{SOC}L_z\sigma_z$
generates large matrix elements between $p_x$- and $p_y$-orbitals, and directly determines the
magnitude of the energy gap at the K-point (Fig. 3c), which is of order $\sim
2\lambda_\text{SOC}$. Second, the $\pi$-bonding sector hybridizes with the
substrate, which breaks inversion symmetry. It features a \textit{Rashba term} which, in
leading order of perturbation theory, couples into the $\sigma$-bonding sector,
yielding a composite effective Hamiltonian:
\begin{equation}
H^{\sigma \sigma}_{\text{eff}}=H_0^{\sigma \sigma}+\lambda_{\text{SOC}}
H_{\text{SOC}}^{\sigma\sigma} + \lambda_{\text{R}}H_{\text{R}}^{\sigma \sigma},
\end{equation}
where $H_0^{\sigma \sigma}$, $H_{\text{SOC}}^{\sigma\sigma}$, and $H_{\text{R}}^{\sigma \sigma}$ are specified in the Supplementary
Information.\\
\indent We derive the prefactors for Bi/SiC as $\lambda_\text{SOC}\sim 0.435\,$eV and
$\lambda_\text{R}\sim 0.032\,$eV. Although the Rashba term $\lambda_\text{R}$ is
much smaller than $\lambda_\text{SOC}$, it induces the large valence band
splitting into states of opposite spin character, while leaving the conduction
band spin-degenerate at K (Fig. 3d). The six valence band maxima are energy-degenerate,
yet carry inverted spin character at the K- versus K'-points, respectively\cite{Xiao.PRL.2007}.
Their large momentum separation and the huge Rashba splitting prevents spin scattering in the
bulk\cite{Xiao.PRL.2012}, as discussed for 2D semiconductors.\\
%
%
\indent The mechanism at work here is very different from the Kane-Mele
mechanism\cite{PhysRevLett.95.226801} for graphene. Its gap emerges due to SOC
between next-nearest neighbours and at the level of second-order perturbation
theory -- and is therefore minute. In other group-IV monolayers, such as
silicene, the SOC remains of next-nearest neighbour
type\cite{PhysRevB.90.085431}, and the relevant orbital is still $p_z$. In
contrast, bismuthene represents the first realized honeycomb material in which
the low-energy physics is driven by the huge \textit{on-site} SOC of $p_x$- and
$p_y$-orbitals. This is also distinguished from HgTe/CdTe quantum wells, where
the small band gap is governed by a complex band situation and opens only upon strain\cite{Bernevig1757,Konig766}.\\
\indent Another fundamental ingredient making the wide-gap QSH phase possible is the
tensile strain regime ($+18\,\%$) of bismuthene on SiC. As argued for
functionalized stanene\cite{PhysRevLett.111.136804}, the topological gap is lost
when external strain shifts another band through the gap. This phase transition
to a trivial system is promoted by compressive strain, while tensile strain
stabilizes the QSH phase. Not surprisingly, compressed
germanene\cite{Zandvliet.PRL.2016} and stanene\cite{SCZhang.NatMat.2015} are
reported as metals, which underpins the importance of the substrate.\\
%
%
\indent A landmark feature of QSH systems are the helical edge channels, connecting
valence and conduction states by two bands of opposite spin (Fig. E5),
irrespective of the edge architecture (zigzag or armchair). Within the bulk gap,
they exhibit an approximately constant density of states (DOS). In a ribbon
simulation, the states rapidly decay towards the bulk within one unit cell, i.e. $\sim 5\,$\AA$\,\,$ (Fig.
E5).\\
\indent Using a tunneling tip, the local DOS (LDOS) is inspected at the atomic scale.
Bismuthene edges exist at SiC substrate steps. Differential tunneling
conductivity (d$I$/d$V$) curves, reflecting the LDOS, have been recorded
along a path towards an uphill step in Fig. 4a. Far from the edge, the spectrum
evidences the large bulk gap of $\sim 0.8\,$eV. Closer to the boundary,
a state emerges filling the entire gap -- as in the ribbon model. Its signal
increases continuously towards the edge, while retaining its shape. A
suppression of d$I$/d$V$ around zero bias might be a tip-related anomaly,
or hint at Luttinger behaviour of the 1D edge states\cite{voit},
which requires further study. The edge DOS in
Fig. 4b shows equivalent behaviour for upper and lower terrace. The d$I$/d$V$
signal (integrated over the gap) in Fig. 4c shows a decay length of $(4\pm
1)\,$\AA, matching well with DFT for simple ribbon edges (Fig. E6).\\
\indent These results bear important differences to a single Bi layer: free-standing it
is hypothesized as QSH phase\cite{PhysRevLett.97.236805} with metallic edge
states\cite{PhysRevB.83.121310}, which is presumed stable for up to several Bi
layers\cite{PhysRevLett.107.136805,PhysRevLett.114.066402}. Edge states for a
single Bi layer on Bi$_2$Te$_3$ of $\sim 2\,$nm extent have been
detected\cite{Yang.PRL.2012} within a small gap of $\sim 70\,$meV, yet with
$E_\text{F}$ in the substrate valence band. In Bi$_{14}$Rh$_3$I$_9$ edge states
are also found, but $E_\text{F}$ is pinned in the bulk conduction
band\cite{Morgenstern.NatPhys.2015}. In contrast, $E_\text{F}$ resides
\textit{within} both the bismuthene gap ($\sim 0.8\,$eV) as well as the SiC gap
($3.2\,$eV), so that conduction is solely governed by the edge states.\\
%
%
\indent While the topological character of the edge states has experimentally yet to be established, e.g.,
by a direct measurement of the QSH effect with its universal
quantized conductance, the remarkable coherence of experimental evidence and
theoretical prediction already makes a strong case for the QSH scenario in
Bi/SiC. For HgTe\cite{Konig766,Roth294} the gap is $\sim 30\,$meV, and in
InAs/GaSb/AlSb quantum wells\cite{PhysRevLett.107.136603} merely $\sim 4\,$meV,
which necessitates experiments below $300\,$mK. The key problem are ''charge
puddles``, where defects push $E_\text{F}$ into the bulk bands, overriding the
1D channel\cite{PhysRevX.3.021003,PhysRevLett.110.216402}. By comparison, the large-gap bismuthene suggests to be operational
at RT. The domain size of $\sim 25\,$nm is expected to be increased by common
technology. The accessible active layer enables many groundbreaking experiments,
such as the response of the edge state to magnetic impurities, or quantum
anomalous Hall setups\cite{Niu.PRB.2015} by chemical functionalization.\\
\subsection{Online Content} Methods, along with any additional Extended Data
display items and Supplementary Information, are available in the online version
of the paper; references unique to these sections appear only in the online
paper.
\newpage
\bibliographystyle{naturemag_noURL}
\bibliography{paper-bic-jul21}



\begin{addendum}
 \item This work was supported by the Deutsche Forschungsgemeinschaft
   (DFG) under Grant SCHA1510/5, the SPP 1666 Priority Program
   ''Topological Insulators``, the DFG Collaborative Research Center SFB
   1170 "ToCoTronics" in W\"urzburg, as well as by the European
   Research Council (ERC) through ERC-StG-Thomale-336012
   ''Topolectrics``. G. L. acknowledges the computing time granted at the Leibniz Supercomputing Centre (LRZ) in Munich. F.R. acknowledges many helpful discussions with M. R. Scholz and J. Aulbach.
	\item[Contributions] J. S. conceived the experiment, S. G.,
F. R., M. B., and J. S. developed the material growth process.
S. G. took the intial STM data. F. R. recorded all STM and STS data shown.   L. D.
conducted the photoelectron spectroscopy. DFT was done
by G. L., and theoretical analysis jointly by G. L. , R. T. and W.
H. All authors contributed to interpreting the data. The
manuscript was written by J. S., R.T., W. H., and R. C. with
suggestions from all other authors.
 \item[Competing Interests] The authors declare no competing financial interests.
 \item[Correspondence] Correspondence and requests for materials should be addressed to J. S. (joerg.schaefer@physik.uni-wuerzburg.de).
\end{addendum}

\newpage

\begin{figure}
\centering
\includegraphics[width=0.95\textwidth]{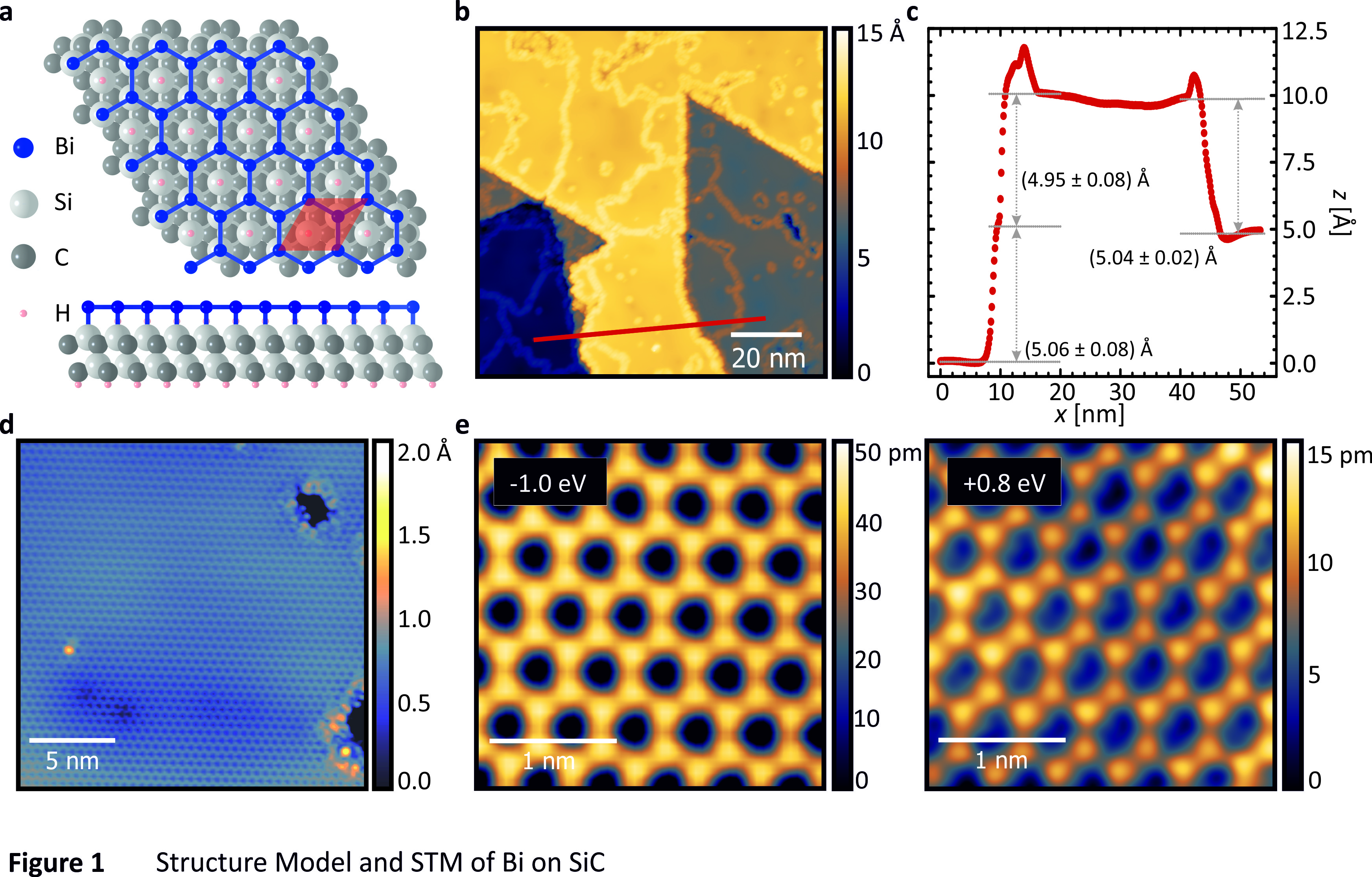} \caption{{\bf Bismuthene on SiC(0001) structural model.} {\bf a},
  Bismuthene layer placed on the threefold-symmetric SiC(0001) substrate in ($\sqrt{3} \times \sqrt{3}$) R$30^{\circ}$ commensurate
registry. \newline
{\bf b}, Topographic STM overview map showing that bismuthene fully covers the substrate. The flakes are of $\sim25$ nm
  extent, limited by domain boundaries. {\bf c}, Substrate step height profile, taken in {\bf b}. The step heights correspond to SiC steps. {\bf d},
The honeycomb pattern is seen on smaller scanframes. {\bf e}, Close-up STM images for empty and occupied states. They confirm the formation of Bi honeycombs.} \label{Fig1}
\end{figure}

\begin{figure}
\centering
\includegraphics[width=1.0\textwidth]{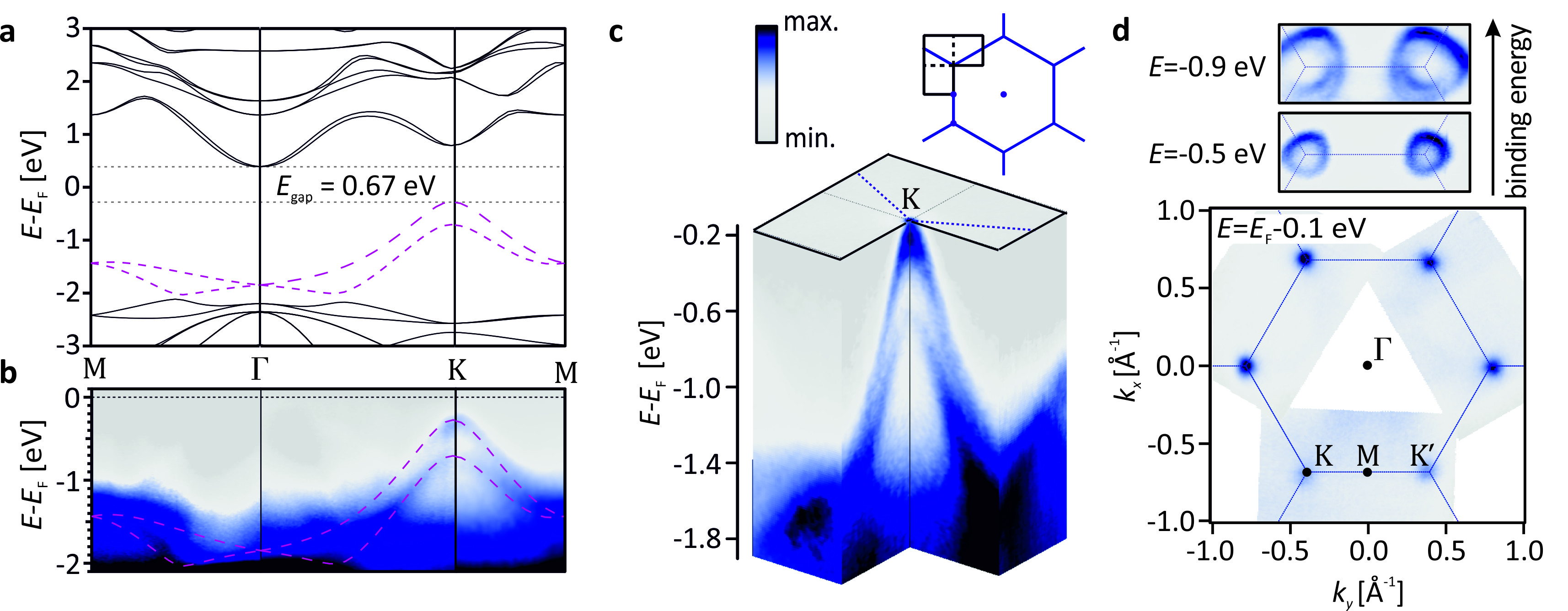}
\caption{{\bf Theoretical band structure and ARPES measurements.}
{\bf  a}, DFT band structure calculation (using a HSE exchange functional)
including SOC, showing the wide band gap and a
significant band splitting in the valence band (dashed line). {\bf b}, ARPES
band dispersion through the Brillouin zone. The band maximum at K
and the valence band splitting are in close agreement with the theoretical
prediction (overlay). The zero of energy ($E_F$) is aligned to the Fermi level of the spectrometer. {\bf c}, Valence band maximum at the K-point with large
SOC-induced splitting in a wide momentum range. {\bf d}, Constant energy
surfaces from ARPES at various binding energies. The cut at low binding energies is taken at the topmost intensity corresponding to the valence band maximum. The six-fold degeneracy of the K- and K'-points of the hexagonal lattice is thereby confirmed.}
	\label{Fig2}
\end{figure}

\begin{figure}
\centering
\includegraphics[width=\textwidth]{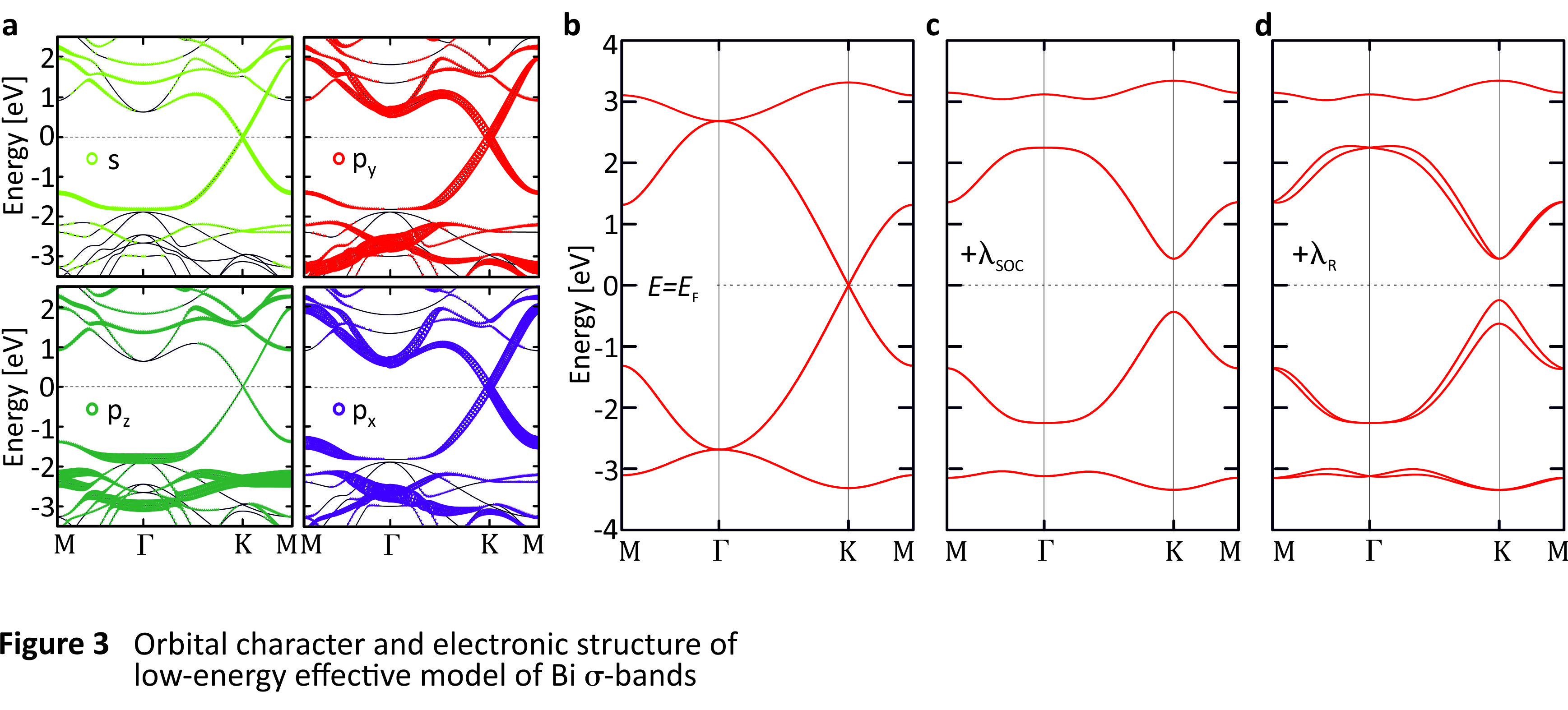}
\caption{{\bf Orbital decomposition (w/o SOC) and the electronic
    structure of the low-energy effective model of Bi $\sigma$-bands.}
{\bf a}, The contribution of  Bi $s$ and $p$-orbitals (without SOC) to the
  electronic structure (w/o SOC) of bismuthene. In each panel, the circle size is proportional to the relative weight of the
  orbital. In Bi/SiC, $p_x$- and $p_y$-orbitals prevail around $E_F$. \newline
	{\bf b}, Electronic structure of the low energy
  effective model without SOC. {\bf c}, Inclusion of the strong atomic SOC opens a huge gap at the K-point. {\bf{d}}, Further including the Rashba term lifts the degeneracy of the topmost valence band, and induces a large splitting with opposite spin character there.}
\label{Fig.3}
\end{figure}

\begin{figure}
\centering
\includegraphics[width=\textwidth]{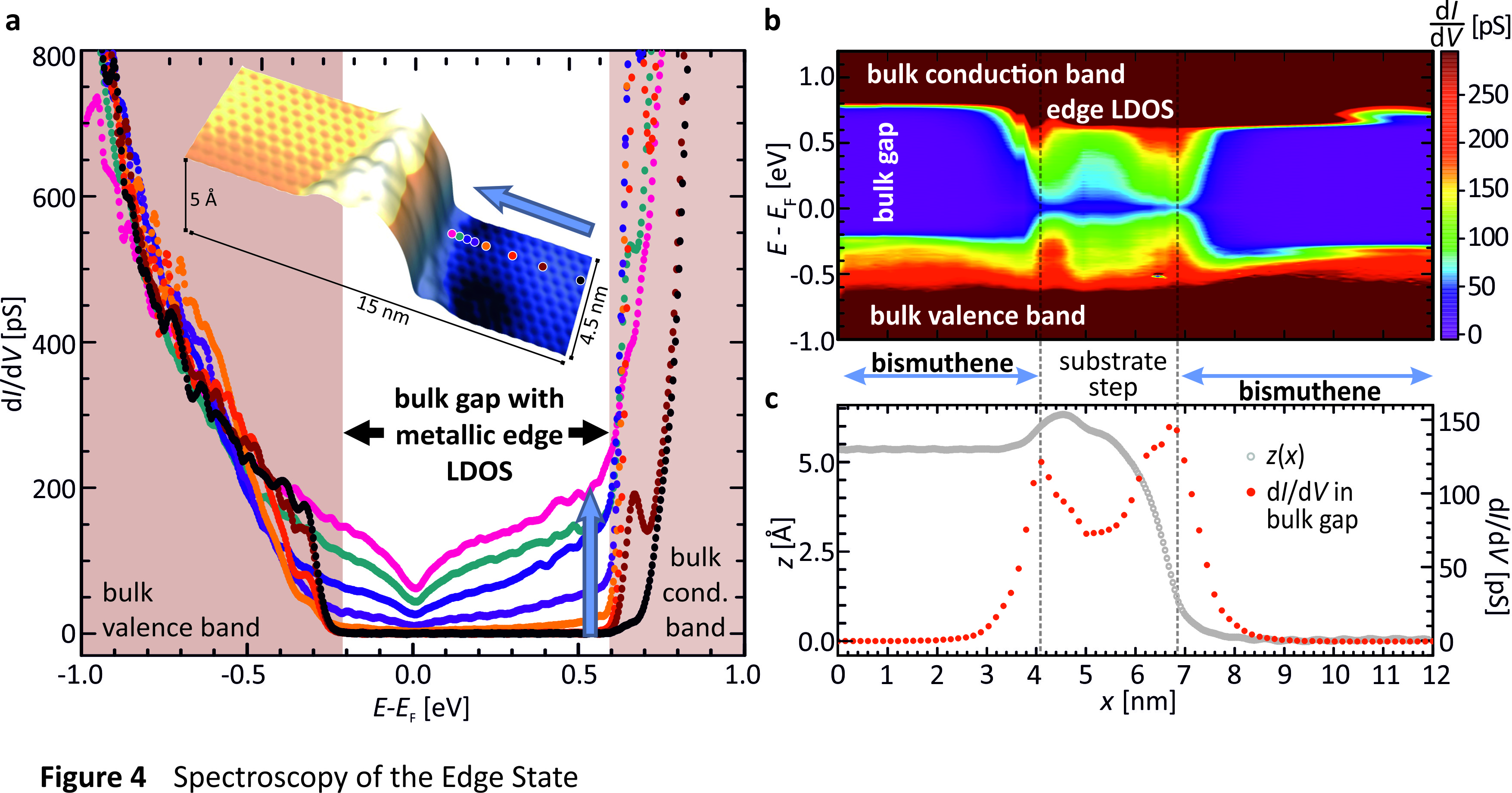}
\caption{{\bf Tunneling spectroscopy of edge states at substrate steps.} {\bf
    a}, Differential conductivity d$I$/d$V$ (reflecting the LDOS) at different distances to the edge. A large gap of $\sim 0.8\,$eV is observed in bulk bismuthene (black curve). Upon approaching the edge, additional signal of increasing strength emerges that fills the entire gap. Inset: STM measurement locations (color-coded dots relate to spectrum color) at uphill substrate step causing the boundary. {\bf b}, Spatially resolved d$I$/d$V$ data across the same step. The d$I$/d$V$ signal of the in-gap states peaks at both film edges (grey dashed lines mark d$I$/d$V$ maxima). {\bf c}, Topographic $z(x)$ line profile of the step , and d$I$/d$V$ signal of bismuthene (integrated over the gap from +0.1 to +0.4$\,$eV), showing an exponential decrease away from the step edge, on either side.}
	\label{Fig.4}
\end{figure}

\end{document}